\documentclass[setspace,amsmath,amssymb]{article}
\usepackage[all]{xy}
\usepackage{graphicx}
\usepackage{setspace}
\usepackage{amsmath}
\usepackage{amssymb}
\usepackage{rsfs}
 
\newcommand{\ra}{\rangle}

\newcommand{\be}{\begin{equation}}
\newcommand{\ee}{\end{equation}}
\newcommand{\bea}{\begin{eqnarray}}
\newcommand{\eea}{\end{eqnarray}}

\begin{document}
\begin{titlepage}

\begin{flushright}
\today
\end{flushright}

\vspace{1in}

\begin{center}

{\bf Quantum measuring systems: considerations from the holographic principle}

\vspace{1in}

\normalsize

{Eiji Konishi\footnote{E-mail address: konishi.eiji.27c@kyoto-u.jp}}

\normalsize
\vspace{.5in}

{\it Graduate School of Human and Environmental Studies,\\
 Kyoto University, Kyoto 606-8501, Japan}
\end{center}

\vspace{1in}

\baselineskip=24pt
\begin{abstract}
In quantum mechanics without application of any superselection rule to the set of the observables, a closed quantum system temporally evolves unitarily, and this Lorentzian regime is characterized by von Neumann entropy of exactly zero.
In the holographic theory in the classicalized ground state, we argue that the unitary real-time evolution of a non-relativistic free particle with complex-valued quantum probability amplitude in this Lorentzian regime can be temporally analytically continued to an imaginary-time classical stochastic process with real-valued conditional probability density in the Euclidean regime, where the von Neumann entropy of the classicalized hologram and the information of a particle trajectory acquired by the classicalized hologram are positive valued.
This argument could shed light on the Euclidean regime of the holographic Universe.

\end{abstract}

\vspace{.1in}


\vspace{.6in}


\end{titlepage}


\section{Introduction}

In the framework of quantum mechanics \cite{Neumann,dEspagnat}, physical phenomena always require a {\it quantum measuring system}, which is the subject of describing them.
In this framework, unitary real-time evolution of the density matrix of a quantum system, which is purely governed by the von Neumann equation, is not substantiated to exist because unitarity means that there are no quantum measuring systems that measure it.
However, in the Universe, the spacetime domain, where the quantum measuring systems exist, is extraordinarily rare.
Does this mean that the particles in the spacetime domain, where no quantum measuring systems exist, are not substantiated to exist?

In this article, we approach this fundamental question in quantum mechanics from the perspective of the {\it holographic principle} \cite{Hol1,Hol2,Hol3}.

For non-holographic theories, three points are relevant to this question:
\begin{enumerate}
\item[(i)] The theories are fundamentally formulated in real time (i.e., in the Lorentzian regime).
\item[(ii)] Non-holographic spacetime is simply the domain, upon which wave functions of particles are defined.
\item[(iii)] In these theories, the state vector of a particle is introduced independently from the concept of space.
\end{enumerate}

In contrast to the non-holographic theories, we invoke the holographic principle \cite{Hol1,Hol2,Hol3,AdSCFT1,AdSCFT2} in the framework of the classicalized holographic tensor network of the ground state of a strongly coupled two-dimensional boundary conformal field theory, which was advocated by the present author in Refs. \cite{EPL1,EPL2,JHAP}.
In the holographic theory, the state vector of a non-relativistic free particle in the bulk spacetime is introduced from the bulk concept of space, which is a highly mixed quantum state after classicalization of the holographic tensor network \cite{JHAP}.
In this theory, we argue that the unitary real-time evolution of a non-relativistic free particle with complex-valued quantum probability amplitude in the Lorentzian regime can be temporally analytically continued to an imaginary-time classical stochastic process with real-valued conditional probability density in the Euclidean regime, where the classicalized hologram, ${\cal H}$, on the boundary spacetime is {\it the} quantum measuring system.
Specifically, in the holographic theory, non-relativistic free particles in the bulk spacetime {\it are always substantiated to exist} as quantum mechanical events occurring with real- or imaginary-time evolution.

The rest of this article is organized as follows.
In Sec. 2, we give preliminaries on the ensemble interpretation of quantum mechanics and the orbital superselection rule.
In Sec. 3, we present a quantitative scheme for real-time evolution of a measured system in the non-holographic theory using the von Neumann entropy and the information gain.
In Sec. 4, we revise this quantitative scheme for a non-relativistic free particle in the holographic theory and resolve the question above.
In Sec. 5, we summarize the overall arguments and conclude the article.

\section{Preliminaries}

\subsection{Ensemble interpretation}

In the ensemble interpretation of quantum mechanics, we reinterpret the probability, which appears in the Born rule in the projective quantum measurement of a discrete observable in a quantum system, as the statistical weight in an ensemble of copies of the quantum system with their state vectors \cite{dEspagnat}.
Here, we use the term {\it ensemble} with the same meaning as the term {\it state}.

In this language, a {\it pure} ensemble means an ensemble of copies of the quantum system with the same state vector, and a non-trivial {\it mixed} ensemble means an ensemble of copies of the quantum system with at least two distinct state vectors.
When we refer to an ensemble as the {\it classical} ensemble with respect to a discrete observable $\widehat{{\cal O}}$, the state vectors in the ensemble are eigenstates of $\widehat{{\cal O}}$.

In the ensemble interpretation, the {\it projective quantum measurement} of a discrete observable $\widehat{{\cal O}}$ in a quantum system consists of two sequential steps: 
\begin{enumerate}
\item[(I)] The dynamical process (called {\it decoherence}) from a given quantum pure ensemble to the classical mixed ensemble with respect to $\widehat{{\cal O}}$.
\item[(II)] The informatical process from the classical mixed ensemble obtained by step (I) to a classical pure ensemble.
\end{enumerate}

In the case of a coherent initial pure ensemble with respect to $\widehat{{\cal O}}$, we refer to step (I) as {\it non-selective measurement} and to step (II) as the subsequent {\it event reading}.\footnote{Intuitively, the term {\it non-selective measurement} means that, after only this dynamical process, events are in a statistical mixture and no particular event is yet singled out.}
The non-selective measurement generates a positive-valued von Neumann entropy, $S_{\rm v.N.}$, of the measured system, and the event-reading system (the quantum measuring system) acquires positive-valued information, $I$.

\subsection{Orbital superselection rule}

The {\it superselection rule} is restriction of the complete set of the observables to the subset of the observables which commute with a given superselection operator \cite{Jauch1,Jauch2}.
In this subsection, we explain the orbital superselection rule, which was introduced by von Neumann in Ref. \cite{Neumann}.

Consider a macroscopic quantum system $A$ (the measurement apparatus) with center of mass orbital observables.

The complete set, ${\cal O}$, of the center of mass orbital observables of system $A$ is generated by the center of mass position variables $\widehat{Q}^a$ ($a=1,\ldots,d$) and the total momentum variables $\widehat{P}^a$ of $A$, which satisfy the canonical commutation relation:
\begin{equation}
[\widehat{Q}^a,\widehat{P}^b]=i\hbar\delta_{ab}\;.
\end{equation}
From this commutation relation, $\widehat{Q}^a$ and $\widehat{P}^a$ cannot be simultaneously measured and have no simultaneous eigenstates.

However, by imposing the orbital superselection rule on the complete set ${\cal O}$, we can redefine $\widehat{Q}^a$ and $\widehat{P}^a$ as $\widehat{{\cal Q}}^a$ and $\widehat{{\cal P}}^a$, respectively, such that
\begin{equation}
\Delta {\cal Q}^a\Delta {\cal P}^a=h\label{eq:OSSR1}
\end{equation}
holds for the constant widths (not the quantum mechanical uncertainties) $\Delta{\cal Q}^a$ and $\Delta {\cal P}^a$ in the discretized spectra of $\widehat{{\cal Q}}^a$ and $\widehat{{\cal P}}^a$, respectively, and 
\begin{equation}
[\widehat{{\cal Q}}^a,\widehat{{\cal P}}^b]=0\label{eq:OSSR2}
\end{equation}
holds.
This commutativity means that $\widehat{{\cal Q}}^a$ and $\widehat{{\cal P}}^b$ can be simultaneously measured and have simultaneous eigenstates.
The validity of this redefinition is attributable to the theorem of the complete orthonormality of the full set of their simultaneous eigenfunctions proved by von Neumann in Ref. \cite{Neumann}.

This orbital superselection rule is attributable to the macroscopicity of the measurement apparatus $A$, which interacts with a microscopic quantum mechanical measured system $S_0$, and the resultant quantum state of $A$ is automatically a classical mixed state of Planck cells $\{h^d\}$ (here, $d$ is the spatial dimensions) in the $\mu$-space of the center of mass of $A$.

Because of the orbital superselection rule giving (\ref{eq:OSSR1}) and (\ref{eq:OSSR2}) of $A$, the non-selective measurement in the combined system $S=S_0+A$ can be performed by the dominant {\it von Neumann-type interaction} \cite{Neumann} between $S_0$ and $A$, and then positive-valued von Neumann entropy $S_{\rm v.N.}$ of $S_0$ is generated.
Namely, the source of information $I$ of $S_0$, which can be acquired by a quantum measuring system $M$, is generated after this non-selective measurement in $S$.

Here, we do not describe the details of the realizable processes of the non-selective measurement in $S$ and the readout of a single event from the mixture of events by $M$.
For these details, see Refs. \cite{JSTAT1,JSTAT2} and Ref. \cite{EPL3}, respectively, and the references therein.

\section{Non-holographic theory}

The von Neumann entropy $S_{\rm v.N.}$ measures the lost information in a quantum mixed state.
This entropy characterizes the {\it unitary real-time evolution} of a quantum system as
\begin{equation}
\delta_t S_{\rm v.N.}=0\label{eq:SvN1}
\end{equation}
for real time $t$ and characterizes the {\it quantum pure state} as
\begin{equation}
S_{\rm v.N.}=0\;.\label{eq:SvN2}
\end{equation}
If the condition (\ref{eq:SvN2}) identically holds temporally, the condition (\ref{eq:SvN1}) follows from it.

In the non-holographic theory, the quantitative schematic of real-time evolution of a measured system $S_0$ is
\begin{equation}
\xymatrix{
\cdots \ar[r]&S_{\rm v.N.}=0\;,\ I=0  \ar[r] & S_{\rm v.N.}>0\;,\ I\ge 0  \ar[r] & S_{\rm v.N.}=0\;,\ I=0  \ar[r] & \cdots}\;.\label{eq:result1}
\end{equation}

As mentioned above, in the presence of the orbital superselection rule due to the macroscopicity of the measurement apparatus $A$, the phase, where the von Neumann entropy of the measured system $S_0$ is positive valued, can appear (i.e., $S_{\rm v.N.}>0$ and $I\ge 0$).

Note that this positive-valued von Neumann entropy of the measured system $S_0$ means that its quantum state is a statistical mixture of quantum mechanical events: the event in the mixture of events is {\it already determined} by the dynamical process, but it is {\it not yet read out} by the quantum measuring system $M$.

On the other hand, the phase where the von Neumann entropy of the measured system $S_0$ is exactly zero (i.e., $S_{\rm v.N.}=0$ and $I=0$) is not substantiated to exist in the framework of quantum mechanics.

\section{Holographic theory: classicalized ground state}

The next argument (\ref{eq:ATN}) in the holographic theory shows the difference between the holographic theory and the non-holographic theory in $d=2$.

By restricting the set of the qubits observables in the boundary conformal field theory (a strongly coupled one-dimensional quantum many-body system at a quantum critical point) from the complete set ${\cal O}$ to the Abelian subset ${\cal A}$ obtained by a superselection rule (here, the superselection operator is the one-qubit third Pauli matrix \cite{JHAP}), the von Neumann entropy of the ground state of the conformal field theory $|\psi\ra_{\rm CFT}$ with a negative Casimir energy \cite{Cardy} in bits is
\begin{equation}
S_{\rm v.N.}[(|\psi\ra_{\rm CFT},{\cal A})]=A_{\rm TN}\label{eq:ATN}
\end{equation}
for the discretized area $A_{\rm TN}$ of the holographic tensor network (the multi-scale entanglement renormalization ansatz \cite{Vidal1,Vidal2}) of the ground state $|\psi\ra_{\rm CFT}$.
(Of course, $S_{\rm v.N.}[(|\psi\ra_{\rm CFT},{\cal O})]=S_{\rm v.N.}[|\psi\ra_{\rm CFT}]$ is zero.)
This formula was derived by the present author in Ref. \cite{EPL2}, and its roots come from the Ryu--Takayanagi formula for the holographic entanglement entropy \cite{RT1,RT2}, which corroborates the holographic tensor network in the bulk space \cite{Swingle1,Swingle2}.
This formula (\ref{eq:ATN}) means that, per site in the classicalized holographic tensor network, there is a locally defined statistical mixture of one {\it spin} degree of freedom with entropy of 1 bit.

From this classical mixed state $(|\psi\ra_{\rm CFT},{\cal A})$, the classical stochastic process of a non-relativistic free particle in the bulk spacetime \cite{Wiener1,Wiener2} is derived in the imaginary time as the readout processes of spin events (i.e., spin eigenstates), which are locally defined at their sites in the classicalized holographic tensor network.

Since the classicalized spin has the action $\hbar$ \cite{EPL2}, there are $I=S_E/\hbar b$ spin events that are read out from the classicalized holographic tensor network for the bit factor $b=\ln 2$ and an off-shell value of the non-relativistic Euclidean action $S_E$ of a free particle in the bulk spacetime.
This argument was made by the present author in Ref. \cite{JHAP}.

In the holographic theory (\ref{eq:ATN}), the quantitative schematic of the time evolution (\ref{eq:result1}) in the non-holographic theory changes drastically for a non-relativistic free particle as
\begin{equation}
{{\xymatrix{
\cdots \ar@{-->}[r]\ar[dr]&S_{\rm v.N.}=0\;,\ I=0  \ar@{-->}[r] \ar@{.>}[d]^{\rm a.c.} & S_{\rm v.N.}>0\;,\ I\ge 0  \ar@{-->}[r] \ar[rd]& S_{\rm v.N.}=0\;,\ I=0  \ar@{-->}[r]\ar@{.>}[d]^{\rm a.c.} & \cdots \\
&S_{\rm v.N.}=A_{\rm TN}\;,\ {\displaystyle{I=\frac{S_E}{\hbar b}}}\ar[ru]&&S_{\rm v.N.}=A_{\rm TN}\;,\ {\displaystyle{I=\frac{S_E}{\hbar b}}}\ar[ru]&}\;.}}
\end{equation}
Here, the Lorentzian phase, where the von Neumann entropy of the particle is exactly zero, is analytically continued to the Euclidean phase, where the von Neumann entropy of the classicalized hologram ${\cal H}$ and the information of a particle trajectory (not spin events) acquired by the classicalized hologram ${\cal H}$ are positive valued.

Here, the {\it analytic continuation} (a.c.) is performed from $t$ (i.e., the Lorentzian regime) to $-i\tau$ (i.e., the Euclidean regime) and means the change
\begin{equation}
\xymatrix{S_{\rm v. N.}=0 \ar@{.>}[r]^{\rm a.c.}&{\displaystyle{I=\frac{S_E}{\hbar b}}}}\;,
\end{equation}
where the complex-valued quantum probability amplitude on the left-hand side is mapped to the real-valued conditional probability density on the right-hand side \cite{JHAP}.

Real time $t$ and imaginary time $\tau$ are independent from (i.e., orthogonal to) each other in the sense that a non-zero real-time increment $\delta t$ in the Lorentzian regime never induces a non-zero imaginary-time increment $\delta \tau$ in the Euclidean regime, and vice versa.
Namely, the choice of a regime is alternative.
However, the unitary real-time evolution of a non-relativistic free particle can be mapped to the imaginary-time classical stochastic process by the Wick rotation, and vice versa \cite{JHAP}.

In the absence of interactions between non-relativistic particles in the bulk spacetime, the additivity of the action in analytical mechanics means that the total information of spin events $I_{\rm tot}$ in bits acquired by the classicalized hologram ${\cal H}$ on the boundary spacetime as the only quantum measuring system in the Euclidean regime is given by the sum of off-shell values of the non-relativistic Euclidean actions $\{S_E^{(n)}\}$ of the free particles in the bulk spacetime divided by $\hbar b$:
\begin{equation}
I_{\rm tot}=\sum_n \frac{S_E^{(n)}}{\hbar b}\;,
\end{equation}
where the index $n$ labels each free particle.

\section{Conclusion}

In quantum mechanics without application of any superselection rule to the set of the observables, a closed quantum system evolves unitarily, and this Lorentzian regime is characterized by von Neumann entropy of exactly zero.

In the non-holographic theory, a quantitative scheme for real-time evolution of a measured system is given by Eq. (\ref{eq:result1}), and the system is not substantiated to exist in this Lorentzian regime.

On the other hand, in the holographic theory in the classicalized ground state, we argued that the unitary real-time evolution of a non-relativistic free particle with complex-valued quantum probability amplitude in this Lorentzian regime can be temporally analytically continued to an imaginary-time classical stochastic process with real-valued conditional probability density in the Euclidean regime, where the von Neumann entropy of the classicalized hologram ${\cal H}$ and the information of a particle trajectory acquired by the classicalized hologram ${\cal H}$ are positive valued.

The overall arguments are summarized in Fig. 1.

\begin{figure}[htbp]
\begin{center}
\includegraphics[width=0.56 \hsize, bb=0 0 260 261]{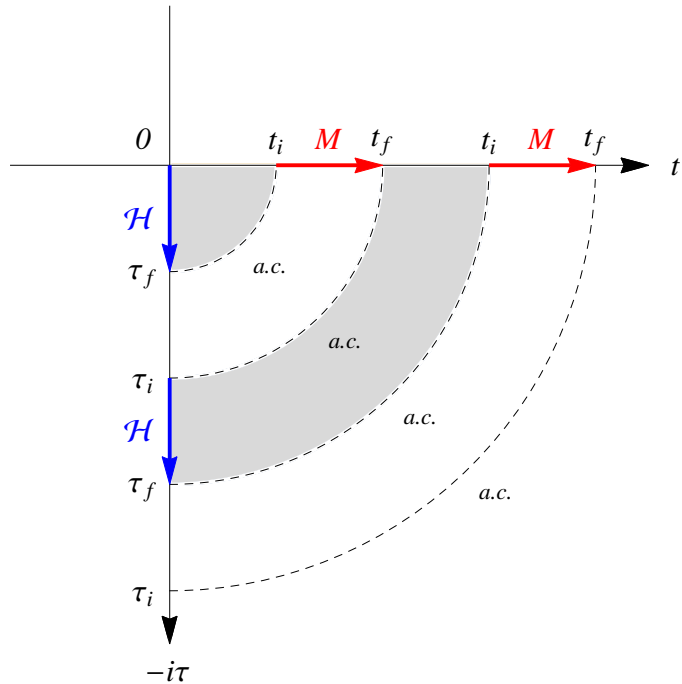}
\end{center}
\caption{Schematic of the quantum measurement processes in the holographic theory in the classicalized ground state.
The blue arrows represent the quantum measurement process by {\it the} quantum measuring system ${\cal H}$ on the boundary spacetime in the Euclidean regime, and the red arrows represent the quantum measurement process by {\it a} quantum measuring system $M$ in the bulk spacetime in the Lorentzian regime.
In the gray domains, the Euclidean regime (a classical stochastic process) and the Lorentzian regime (unitary quantum mechanics) of a non-relativistic free particle in the bulk spacetime are dual to each other by the Wick rotation.
During $t_i\le t \le t_f$, one or more non-unitary quantum measurements are performed by a quantum measuring system $M$ in the bulk spacetime in the Lorentzian regime, and no dual Euclidean regime exists.}
\end{figure}

In conclusion, in the holographic theory in the classicalized ground state, a non-relativistic free particle in the bulk spacetime is always substantiated to exist as quantum mechanical events occurring with real- or imaginary-time evolution by the quantum measuring systems; in the Euclidean regime, the classicalized hologram ${\cal H}$ on the boundary spacetime is the only quantum measuring system.


\begin{thebibliography}{99}
\bibitem{Neumann}J. von Neumann, {\it Mathematical Foundations of Quantum Mechanics} (Princeton University Press, Princeton, NJ, 1955).
\bibitem{dEspagnat}B. d'Espagnat, {\it Conceptual Foundations of Quantum Mechanics}, 2nd edn. (W. A. Benjamin, Reading, Massachusetts, 1976).
\bibitem{Hol1}G. 't Hooft, arXiv:gr-qc/9310026.
\bibitem{Hol2}L. Susskind,
J. Math. Phys. {\bf 36}, 6377 (1995).
\bibitem{Hol3}R. Bousso,
Rev. Mod. Phys. {\bf 74}, 825 (2002).
\bibitem{AdSCFT1}J. M. Maldacena, 
Adv. Theor. Math. Phys. {\bf 2}, 231 (1998).
\bibitem{AdSCFT2}O. Aharony, S. S. Gubser, J. M. Maldacena, H. Ooguri and Y. Oz,
 Phys. Rep. {\bf 323}, 183 (2000).
\bibitem{EPL1}E. Konishi, 
EPL {\bf 129}, 11006 (2020).
\bibitem{EPL2}E. Konishi,
 EPL {\bf 132}, 59901 (2020), arXiv:1903.11244 [quant-ph].
\bibitem{JHAP}E. Konishi,
JHAP {\bf 1}, (1) 47-56 (2021).
\bibitem{Jauch1}J. M. Jauch, 
Helv. Phys. Acta. {\bf 33}, 711 (1960).
\bibitem{Jauch2}J. M. Jauch and B. Misra,
Helv. Phys. Acta. {\bf 34}, 699 (1961).
\bibitem{JSTAT1}E. Konishi, J. Stat. Mech. 063403 (2018).
\bibitem{JSTAT2}E. Konishi, J. Stat. Mech. 019501 (2019).
\bibitem{EPL3}E. Konishi,
EPL {\bf 136}, 10004 (2021).
\bibitem{Cardy}J. L. Cardy and I. Peschel,
Nucl. Phys. B {\bf 300}, 377 (1988).
\bibitem{Vidal1}G. Vidal, 
Phys. Rev. Lett. {\bf 99}, 220405 (2007).
\bibitem{Vidal2}G. Vidal, 
Phys. Rev. Lett. {\bf 101}, 110501 (2008).
\bibitem{RT1}S. Ryu and T. Takayanagi, 
Phys. Rev. Lett. {\bf 96}, 181602 (2006).
\bibitem{RT2}M. Rangamani and T. Takayanagi, 
Lect. Notes Phys. {\bf 931}, 1 (2017).
\bibitem{Swingle1}B. Swingle, 
Phys. Rev. D {\bf 86}, 065007 (2012).
\bibitem{Swingle2}B. Swingle,
Annu. Rev. Condens. Matter Phys. {\bf 9}, 345 (2018).
\bibitem{Wiener1}N. Wiener,
J. Math. and Phys. {\bf 2}, 131 (1923).
\bibitem{Wiener2}N. Wiener,
Proc. London Math. Soc. Ser. 2 {\bf 22}, 454 (1924).

\end{thebibliography}
\end{document}